\ifdefined\isabridged
\documentclass[journal,compsoc,final]{IEEEtran}
\else
\documentclass[sigconf]{acmart}
\usepackage{lmodern}
\fi
\usepackage[T1]{fontenc}
\usepackage[utf8]{inputenc}
\usepackage{letltxmacro}
\usepackage{todonotes}
\usepackage{xspace}
\usepackage{hyperref}
\usepackage{colortbl}
\usepackage{pifont}
\usepackage{ulem}
\usepackage{booktabs}
\usepackage[font=small]{caption}
\usepackage{multirow}
\usepackage{tablefootnote}
\usepackage{enumerate,mdwlist}
\usepackage{paralist}

\usepackage{url}

\newif\ifabridged
\newif\ifnotabridged
\newif\ifanonymous
\newif\ifnotanonymous

\ifdefined\isabridged
\abridgedtrue
\fi

\ifabridged\notabridgedfalse
\else\notabridgedtrue
\fi

\ifdefined\isanonymous
\anonymoustrue
\fi

\ifanonymous\notanonymousfalse
\else\notanonymoustrue
\fi

\LetLtxMacro{\todonote}{\todo}
\renewcommand{\todo}[2][]
{\todonote[inline, caption={#2}, size=\footnotesize, #1]
{\renewcommand{\baselinestretch}{0.5}\selectfont#2\par}}

\newcommand\titlefootnote[1]{  \begingroup
  \renewcommand\thefootnote{}\footnote{#1}  \addtocounter{footnote}{-1}  \endgroup
}

\newcommand{\dOne}{\ding{182}}
\newcommand{\dTwo}{\ding{183}}
\newcommand{\dThree}{\ding{184}}
\newcommand{\dFour}{\ding{185}}
\newcommand{\dFive}{\ding{186}}
\newcommand{\dSix}{\ding{187}}
\newcommand{\dSeven}{\ding{188}}

\newcommand{\dwOne}{\ding{192}}
\newcommand{\dwTwo}{\ding{193}}
\newcommand{\dwThree}{\ding{194}}

\newcommand{\dc}{Controller\xspace}
\newcommand{\oem}{\ensuremath{\mathit{OEM}}\xspace}
\newcommand{\oems}{\ensuremath{\mathit{OEMs}}\xspace}
\newcommand{\sd}{Distributor\xspace}
\newcommand{\dev}{Device\xspace}

\newcommand{\hydra}{{\sf\small HYDRA}\xspace}
\newcommand{\attkey}{$K_{\mathit{Att}}$\xspace}
\newcommand{\oemkey}{$K_{\mathit{OEM}}$\xspace}
\newcommand{\patt}{\emph{PR}$_{Att}$\xspace}

\newcommand{\acro}{{\sf\small ASSURED}\xspace}
\newcommand{\acrotitle}{{\sf ASSURED}\xspace}

\newcommand{\dtoprule}{\specialrule{.8pt}{0pt}{1pt}            \specialrule{.8pt}{0pt}{\belowrulesep}}

\usepackage{lipsum}
\newcounter{question}
\ifabridged
\makeatletter
\def\ps@IEEEtitlepagestyle{
  \def\@oddfoot{\myfooter}
  \def\@evenfoot{}
}
\def\myfooter{
  {\footnotesize
  \begin{minipage}{\textwidth}
  \centering
  This article was presented in the International Conference on Embedded Software 2018 and appears as part of the ESWEEK-TCAD special issue.
  \end{minipage}
  }
}
\else
\renewcommand\footnotetextcopyrightpermission[1]{} %
\fi

\title{ASSURED: Architecture for Secure Software Update of Realistic Embedded Devices}

\ifnotanonymous
\ifabridged  %
\author{
  \IEEEauthorblockN{N.~Asokan~\IEEEmembership{Fellow,~IEEE},
                    Thomas~Nyman,
                    Norrathep~Rattanavipanon,
                    Ahmad-Reza~Sadeghi,
                    and~Gene Tsudik~\IEEEmembership{Fellow,~IEEE}} \\
\IEEEcompsocitemizethanks{%
  \IEEEcompsocthanksitem Author names are listed in alphabetical order.
  \IEEEcompsocthanksitem N.~Asokan and T.~Nyman are with the Secure Systems
  Group of Aalto University, Espoo 02150, Finland. Portions of this research were done while T. Nyman was with Trustonic Oy, 
  Helsinki, Finland. E-mail:  asokan@acm.org, thomas.nyman@aalto.fi.
  \IEEEcompsocthanksitem N.~Rattanavipanon and G.~Tsudik are with the Computer Science Department, University of California, Irvine, CA 92697-3435, USA. 
  E-mail: \{nrattana,gts\}@uci.edu.
  \IEEEcompsocthanksitem A-R. Sadeghi is with the Technische Universit\"at Darmstadt, Darmstadt 64289, Germany. 
  E-mail: ahmad.sadeghi@trust.tu-darmstadt.de.
}}
\else
\author{N. Asokan}
\authornote{Author names are listed in alphabetical order.}
\affiliation{%
  \institution{Aalto University, Finland}
}
\email{asokan@acm.org}

\author{Thomas Nyman}
\authornote{Portions of this research were done while the author was with Trustonic Oy, Finland}
\affiliation{%
  \institution{Aalto University, Finland}
}
\email{thomas.nyman@aalto.fi}

\author{Norrathep Rattanavipanon}
\affiliation{%
  \institution{University of California, Irvine, USA}
}
\email{nrattana@uci.edu}

\author{Ahmad-Reza Sadeghi}
\affiliation{%
  \institution{Technische Universit\"at Darmstadt, Germany}
}
\email{ahmad.sadeghi@trust.tu-darmstadt.de}

\author{Gene Tsudik}
\affiliation{%
  \institution{University of California, Irvince, USA}
}
\email{gts@uci.edu}
\fi

\begin{document}

\ifabridged
\IEEEtitleabstractindextext{%
\begin{abstract}

Secure firmware update is an important stage in the IoT device life-cycle.
Prior techniques, designed for other computational settings, are not readily suitable for
IoT devices, since they do not consider idiosyncrasies of a realistic large-scale IoT deployment. 
This motivates our design of \acro, a secure and scalable update framework for IoT.
\acro includes all stakeholders in a typical IoT update ecosystem, while providing end-to-end
security between manufacturers and devices. To demonstrate its feasibility and practicality, \acro
is instantiated and experimentally evaluated on two commodity hardware platforms. Results show 
that \acro is considerably  faster than current update mechanisms in realistic settings.

\end{abstract}

}
\else
\begin{abstract}

Secure firmware update is an important stage in the IoT device life-cycle.
Prior techniques, designed for other computational settings, are not readily suitable for
IoT devices, since they do not consider idiosyncrasies of a realistic large-scale IoT deployment. 
This motivates our design of \acro, a secure and scalable update framework for IoT.
\acro includes all stakeholders in a typical IoT update ecosystem, while providing end-to-end
security between manufacturers and devices. To demonstrate its feasibility and practicality, \acro
is instantiated and experimentally evaluated on two commodity hardware platforms. Results show 
that \acro is considerably  faster than current update mechanisms in realistic settings.

\end{abstract}

\fi

\maketitle

\ifnotabridged

\titlefootnote{
\emph{International Conference on Embedded Software (EMSOFT'18),\\ October 2018, Turin, Italy} \copyright\ 2018 IEEE\\
  This is the author's version of the work. It is posted here for
  your personal use. Not for redistribution. The definitive Version
  of Record was published in
  \emph{IEEE Transactions on Computer-Aided Design of Integrated Circuits and Systems, vol. 37, no. 11, Nov. 2018.}\\
  \url{https://doi.org/10.1109/TCAD.2018.2858422}}
\fi

\section{Introduction}
Deploying insecure Internet-of-Things (IoT) devices can have disastrous consequences, as demonstrated
by large-scale IoT botnets, such as Mirai~\cite{Antonakakis17} and 
Reaper\footnote{\url{https://www.arbornetworks.com/blog/asert/reaper-madness/}}. IoT devices are ideal 
malware targets, for several reasons: First, Internet-connected devices are inherently more exposed to 
remote exploitation. Second, embedded systems are notoriously difficult to update, which often leaves
known vulnerabilities unpatched. Third, many such devices operate in a mostly unattended fashion, 
which means that timely discovery of compromise is unlikely.

Once an IoT device is deployed, the ability to remotely update its device's firmware is critical to maintaining security 
over its lifetime. In many real-world scenarios, devices must be deployed in remote or 
inaccessible locations, rendering physical (manual) maintenance impossible or prohibitively 
expensive. Remote \emph{"Over-the-Air"} (OTA) delivery of firmware updates allows manufacturers 
to deliver new features or functionality, as well as to patch bugs and flaws. However, if designed poorly, 
insecure update mechanisms may be exploited by the adversary, causing victim devices to malfunction, 
cease operation, or fall under adversarial control. 
Some prior update techniques (geared for different settings) meet security requirements 
under specific assumptions. For example, TUF~\cite{samuel2010survivable} is an update delivery framework  
resilient to key compromise, while its descendant Uptane~\cite{karthikuptane} extends and adapts 
TUF to support secure updates for automotive systems. 
However, both techniques requires direct interaction between the manufacturer and the devices
in order to specify device-specific constraints on the update process.
This makes them unsuitable for large-scale 
IoT deployments, where updates may be delivered via broadcast, or from third-party \emph{Content 
Delivery Networks} (CDNs). Hence, the update mechanism can not rely on interactive protocols or on
transport-level security. Also, TUF and Uptane do not support 
verification of proper update installation on target devices.  

Some proposals for secure firmware updates on resource-constrained devices 
(e.g., SCUBA~\cite{seshadri2006scuba} and PoSE~\cite{perito2010secure})
allow the updater to obtain a verifiable proof  of successful update. However, they involve
strong assumptions (e.g., an optimal checksum function or strictly local
communication),  or lack support for
update robustness, i.e., roll-back to a previous firmware version if the current update fails.
Such issues make them unsuitable for realistic IoT deployments. This is discussed further in Section
\ref{sec:background} which overviews related work.

\noindent\textbf{Goals and Contributions:} \begin{itemize}[$\bullet$]
\item \textit{IoT Update Ecosystem:} We identify essential roles in the IoT secure software update ecosystem and show that they 
cannot be directly incorporated into state-of-the-art secure update methods~\cite{samuel2010survivable,karthikuptane}. 
We also identify objectives for an IoT secure software update system. (Section~\ref{sec:model})
\item \textit{Secure Firmware Update Framework:} We propose \acro, which: (a) provides end-to-end security 
by combining an existing reliable update delivery framework (e.g., TUF) with an authorization mechanism  
that allows manufacturers to specify update constraints, and (b) allows a local authority
to specify constraints for -- and verify -- successful update deployment. (Section~\ref{sec:design})
\item \textit{Realization \& Evaluation:} We instantiate \acro on two low-end security architectures: 
\hydra~\cite{hydra} and  ARM TrustZone-M~\cite{ARM-TrustZone-M}. We also demonstrate its practicality via proof-of-concept implementations on two commodity platforms:
I.MX6-SabreLite~\cite{sabre} (Section~\ref{sec:assured-hydra}) and ARM Cortex-M23 microcontroller 
prototyping system equipped with TrustZone-M (Section~\ref{sec:assured-tzm}). 
Our evaluation shows that \acro improves upon current update architectures in terms of deployability and performance in realistic IoT settings. It also meets the objectives we identified for different stakeholders in the IoT update ecosystem.

\end{itemize}

\section{Background \& Prior Work}\label{sec:background}
This section overviews several related topics. 
\subsection{Boot Integrity}
\label{sec:boot-integrity}
Platform boot integrity is a fundamental requirement for any system designed to resist copying, corruption, or compromise. 
\emph{Secure} or \emph{authenticated boot}~\cite{Pearson02} mechanisms examine integrity of the system's software components 
at boot time, thus detecting changes to the system's trusted state. 

In {\em secure boot}, each step in the boot process verifies a public key
signature on the next step in the boot chain, before it is launched. The source of trust in the secure boot process typically originates from a 
\emph{Static Root-of-Trust}, such as an immutable piece of code and a private key, imprinted (hard-coded) by the device manufacturer. 
A software image must be signed by its manufacturer before deployment, making it impractical to verify configuration information 
provided by the system administrator that controls the device during its operation. 

In {\em authenticated boot}, each step of the boot process is measured, e.g., by computing a cryptographic hash over the software image 
and platform configuration information; the resulting measurement is stored in a way that allows it to be securely retrieved later. Unlike 
secure boot, authenticated boot permits any software component to run. However, the securely stored measurement can be used for 
local access control decisions (e.g., access to hardware-based keys), or for producing a signed statement of the system's state to a 
remote verifier, as described in Section~\ref{sec:remote-attestation}. Authenticated boot relies on a Root-of-Trust for guaranteeing 
unforgeability of measurements.

Secure or authenticated boot are standard features in modern PC~\cite{Paul15} and mobile platforms~\cite{Asokan14}, although their 
architectural realizations can differ significantly across platforms. Boot integrity is also important for embedded platforms, where its
use has been mainly to protect against memory corruption~\cite{Donovan15}. For instance, virtually all microcontroller units (MCUs) check operating 
integrity at initialization, or during recovery from a low-voltage condition, e.g., by computing a \emph{Cyclic Redundancy Checksum} 
(CRC) of the software image, and comparing it with a CRC stored in persistent storage, typically flash memory. However, CRC-based 
checks do not defend against attacks on device's boot integrity, since an attacker who modifies the code on the device can bypass 
the CRC check via specially crafted software images, or even by modifying the reference CRC in flash. Therefore, modern MCU platforms 
employ cryptographic hash algorithms (instead of CRCs) and one-time-programmable fuses to store reference measurements used for 
secure boot. The use of cryptographic algorithms for secure boot and for communication in resource-constrained MCUs triggers 
inclusion of cryptographic hardware accelerators, even in very small MCUs. 
\subsection{Remote Attestation}
\label{sec:remote-attestation}
Remote attestation is a process whereby a trusted entity (verifier) remotely measures internal state of a untrusted and 
possible compromised device (prover), in order to determine whether the latter is in a benign state. 
Current remote attestation approaches can be partitioned into three groups: hardware-based, software-based and hybrid.
Hardware-based attestation relies on security provided by dedicated hardware features such as a 
Trusted Platform Module (TPM)~\cite{tpm} or Intel's SGX~\cite{costan2016intel}. Such hardware features are generally 
not viable for resource constrained IoT devices, such as MCUs, due to their complexity and cost.

On the other hand, software-based attestation requires no hardware features at all. Instead, it
assumes: (1) consistent timing characteristics of the measurement process on the prover,  (2) existence of an 
optimal (space- and time-wise) checksum function~\cite{seshadri2005pioneer, li2011viper, seshadri2004swatt}. 
Unfortunately, these assumptions only hold when attestation is performed over one-hop communication, along
with an idealized checksum function. Consequently, software-based methods are unsuitable for remote attestation
in realistic settings, e.g., over the Internet.

Hybrid remote attestation is exemplified by SMART~\cite{eldefrawy2012smart} architecture. It imposes minimal changes to existing MCUs.
SMART requires immutability of attestation code and key by storing them in a read-only memory region. 
SMART also utilizes hardwired MCU access control rules to ensure that: (1) access to the attestation key is restricted to 
attestation code, and (2) execution of attestation code is atomic, i.e., uninterruptible and executed as a whole. 
A follow-on result, TrustLite~\cite{koeberl2014trustlite}, provides a more flexible way to specify these access control rules.
Access control configuration in TrustLite can be programmed in software at compile time and enforced by an additional feature, 
EA-MPU: Execution-Aware Memory Protection Unit. Unlike SMART, TrustLite does not require uninterruptible execution of
attestation code, since its CPU Exception Engine is modified to support secure interrupt handling. A subsequent result, 
TyTan~\cite{brasser2015tytan}, extends TrustLite to support dynamic access control configuration and real-time guarantees.

\begin{figure*}[t]
  \centering
    \ifabridged
    \includegraphics[width=\hsize]{figures/generic-architecture.pdf}
    \else
    \includegraphics[width=\hsize]{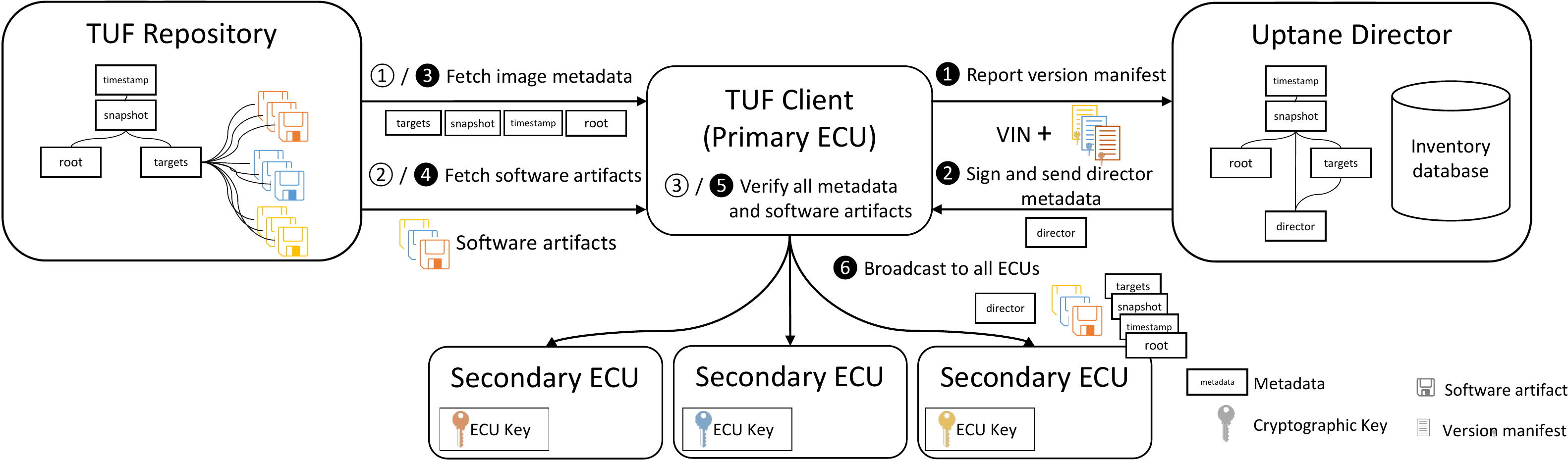}
    \fi
    \caption{Sequence of update distribution events in TUF and Uptane. \dwOne\ to \dwThree\ depict the update process for TUF, 
    while \dOne\ to \dSix\ depict the extended update process employed by Uptane.}
  \label{fig:tuf-uptane}
\end{figure*}

\subsection{Secure Updates}
\noindent\textbf{The Update Framework (TUF)}\cite{samuel2010survivable} is a generic security framework designed 
to integrate with existing software repositories. TUF adds a new layer of signed metadata, including file sizes and cryptographic 
hashes of file content. Figure~\ref{fig:tuf-uptane} (\dwOne\ through \dwThree) illustrates the sequence of events in 
TUF-based update distribution. TUF clients (end-hosts) periodically poll repositories for changes and fetch this metadata 
(\dwOne), and new software artifacts as needed (\dwTwo). By verifying the metadata (\dwThree), clients detect whether 
files or metadata have been manipulated.

TUF assigns responsibility of signing different parts of metadata to different \emph{roles}. In order to improve 
resilience against key compromise, all roles can use one or more distinct key-pairs and require clients to 
validate a threshold number of signatures of the role's keys. TUF defines four fundamental roles necessary 
to meet its security goals: \emph{root},  \emph{targets}, \emph{snapshots}, and \emph{timestamp}.

{\em The root role} acts as a Certification Authority (CA) for the repository. It signs public keys of all other top-level roles. 
TUF clients must receive the root role's trusted public keys out-of-band, e.g., at manufacture or install time. Since 
the root role's keys act as roots-of-trust in TUF, they should be stored offline, physically disconnected from the 
Internet to minimize the risk of compromise. 

{\em The targets} role signs metadata describing software artifacts which can be trusted by clients. Since software 
may be originated by different sources, the targets role may delegate full or partial trust to an auxiliary role with a 
separate set of key-pairs. Partial delegation limits the set of files that the role is allowed to indicate as trusted. 
A role with delegated trust can delegate this ability further. 

{\em The snapshots role} signs metadata that confirms the latest version of all other TUF metadata stored in 
the repository, except the timestamp role metadata described below. 

All TUF metadata is associated with an expiration time. In addition, {\em the timestamp role} periodically 
signs a statement indicating the latest version of the snapshot metadata even if there have been no updates.

\noindent\textbf{Uptane}~\cite{karthikuptane} is an instantiation of TUF customized for software distribution to 
automotive systems. In order to manage updates to numerous and diverse computerized components 
found in modern vehicles, Uptane extends TUF with an additional \emph{director} repository. This repository allows 
an \emph{Original Equipment Manufacturer} (OEM) more control of software images deployed in individual ECUs.

Figure~\ref{fig:tuf-uptane} (\dOne\ through \dSix) illustrates the sequence of events in Uptane-based update distribution. 
Uptane designates one Engine Control Unit (ECU) of each vehicle as \emph{primary}; it orchestrates delivery of updates between 
the repository and \emph{secondary} ECUs. As part of the update process, the primary ECU reports a 
\emph{vehicle version manifest} (containing a signed statement from each ECU about its software configuration) 
to the remote director repository, along with the \emph{Vehicle Identification Number} (VIN) (\dOne). The director repository  
determines the correct, up-to-date software configuration for each ECU in the vehicle identified by the VIN. It also 
signs director metadata that contains instructions bound to the unique serial number of each ECU, 
describing all software artifacts each ECU must install (\dTwo). The primary ECU fetches and verifies all metadata 
and software artifacts on behalf of secondary ECUs (\dThree\ through \dFive) and distributes them over the vehicle's 
local-area network (\dSix). The metadata is broadcast to all ECUs. 

To protect against compromise of the primary ECU or man-in-the-middle attacks (MiTM) attack originating 
within the vehicle's internal network, each secondary ECU re-verifies the metadata and software artifacts it receives 
from the primary ECU. However, each secondary ECU might not be equipped to fully verify or store all repository 
metadata. To accommodate such ECUs, Uptane relaxes verification requirements for \emph{partial verification} 
ECUs, which only receive and verify director's metadata and software artifacts specified therein for that ECU.

\subsection{Secure Update via Remote Attestation}
Several prior secure update techniques use pre-existing remote attestation designs. For example, 
\emph{SCUBA}~\cite{seshadri2006scuba} can be used to repair a compromised sensor 
through firmware updates. SCUBA utilizes an authentication mechanism and software-based attestation to identify 
memory regions infected by malware and transmits the repair update to replace these regions.
However, the attestation technique based on self-checksumming code heavily relies on
consistent timing characteristics of the measurement process the use of an optimal checksum function. 
Due to these assumptions, SCUBA is not a suitable approach for IoT settings~\cite{castelluccia2009difficulty}.

Perito and Tsudik~\cite{perito2010secure} present a simple secure firmware update technique using so-called
Proofs of Secure Erasure (PoSE-s). The idea is for prover to perform secure erasure before retrieving and 
installing a new update, from clean slate.  Secure erasure is achieved by filling all of prover's memory with 
(uncompressable) randomness chosen by verifier. Prover then returns a snapshot of the new memory contents
to verifier as a proof of secure erasure. This proof guarantees that prover is now in a benign clean state and ready to perform 
an update. Subsequent work~\cite{karame2015secure, karvelas2014efficient} improved the original method by 
reducing time, energy and bandwidth overheads.  However, all PoSE-based update techniques work only 
in a one-hop prover/verifier setting. Furthermore,
they do not support {\em robust updates}, i.e. ability to retain previous version(s), in case a roll-back is
required, e.g., if the current update cannot be successfully completed. Whereas, in \acro, prover
can isolate untrusted software, thus secure erasure is not needed and update robustness can be achieved.

\subsection{ARM TrustZone}
\label{sec:trustzone}
ARM microprocessors are RISC-based low-power processors that can be found in many modern devices, 
such as smartphones, smart TVs, smart watches, tablets and various other computing devices. \textit{Cortex-A} 
series application processors are commonly deployed in mobile devices, networking equipment and other home 
and consumer devices. \textit{Cortex-M} series embedded processors are used in MCUs that require low cost
and energy efficiency, such as sensors, wearables and small robotic devices. 

Since 2003, ARM application processors have featured \emph{TrustZone} Security Extensions~\cite{ARM-TrustZone}, 
a hardware feature that aims to reduce attack surface of security-critical code by separating processor operation 
into two distinct states: \emph{Secure} and \emph{Non-secure}. In Non-secure state, hardware-based access control 
prevents less trusted software (OS and applications) from accessing resources belonging to the Secure state. 
The ARMv8-M architecture also supports TrustZone Security Extensions in Cortex-M cores. Guarantees provided by 
TrustZone in Cortex-M processor are similar to that by TrustZone in Cortex-A~\cite{ARM-TrustZone-M}, although the 
microarchitectural realizations of TrustZone Security Extensions differ significantly between Cortex-A and Cortex-M processors.

ARM-based IoT devices commonly utilize either low-end Cortex-A processors ( $<1$GHz cores typically \emph{without} TrustZone) 
or Cortex-M microcontrollers, ranging from few tens to a few hundred MHz. The latest additions to the Cortex-M processor family:
Cortex-M23\footnote{\scriptsize\url{https://developer.arm.com/products/processors/cortex-m/cortex-m23}} and 
Cortex-M33\footnote{\scriptsize \url{https://developer.arm.com/products/processors/cortex-m/cortex-m33}}, also feature TrustZone Security Extensions.

\section{System Model}
\label{sec:model}
In this section, we identify essential stakeholders in the IoT firmware update ecosystem.
We then specify anticipated adversarial capabilities and assumptions.
Next, we describe the requirements for \acro and discuss how to realize them on a low-end device.

\subsection{Stakeholders}
\label{sec:stakeholders}
We adopt the same stakeholder model as the one used in the Software Updates for Internet of 
Things (SUIT) Working Group\footnote{\url{https://datatracker.ietf.org/wg/suit/about/}} of the Internet Engineering 
Task Force (IETF)~\cite{Moran-FUD1}. It includes four types of stakeholders:
\begin{itemize}[$\bullet$]
  \item \textbf{Original Equipment Manufacturer (\oem)}. 
  Produces devices, issues the initial firmware and releases subsequent updates. During manufacturing, \oem can securely install 
  cryptographic keys on its devices~\footnote{While software itself may be produced by a third-party developer, we assume that OEM 
  always controls its distribution. While not inconceivable, we are unaware of any cases of a software 
  developer directly distributing firmware and/or its updates to IoT devices from multiple OEMs.}
  \item \textbf{(Software) Distributor}. 
      Essentially plays the role of a surrogate in the update distribution process.
  Since \oems may not wish to build and maintain  the complex infrastructure to support logistics of 
  software distribution at scale, update distribution can be outsourced to software distributors, such as CDNs.
\item \textbf{(Domain) Controller}. 
  Responsible for the upkeep, configuration, and reliable operation of 
  devices within its administrative domain. The domain may be defined by physical proximity, e.g., 
  devices in the vicinity that are reachable from \dc via local connectivity, e.g., WiFi or Bluetooth Low Energy. 
      Alternatively, the domain that \dc is responsible for may be defined organizationally, e.g., in cases where \dc is operated as a cloud-hosted service.
\item \textbf{(Connected) Device}. 
  The ultimate target of updates. We focus on resource-constrained (low-end) connected 
  devices, such as: sensors, actuators, hybrids of both, or any other embedded devices that 
  operate under strict resource limitations in terms of memory, storage and processing power. 
\end{itemize}
The \oem and \dc must be able to specify constraints on updates to be deployed on \dev.  For instance, \oem might use 
the same signing key to sign updates for different device variations. Thus, it needs to ensure that a particular device only 
installs updates for the correct variation.
A software update might also be issued to enable or disable a feature for a particular subset of devices, e.g., enable debug 
for a development device.
\dc may wish to constrain update deployment time, i.e., during a regular maintenance window, or when a device is otherwise idle. 
Updates may also be deployed as differential patches, i.e., updates do not contain full software images, but only the changes between the previous software version, and the updated version. In such cases the \oem must be able to place constraints on the order in which the updates are installed to ensure that each patch applies cleanly, and the resulting software configuration in \dev remains consistent at all times. 

\subsection{Adversary Model}
\label{sec:adv-model}
We base our adversarial model on the subset of realistic attack types enumerated in \cite{abera2016things}:
\begin{enumerate}[I]
  \item\label{sc:adv-model:ra} \textit{Remote Adversary} can compromise untrusted file servers or 
  cloud storage infrastructure components that store firmware updates before they reach \dev. 
  Remote adversary may also attempt to remotely exploit software vulnerabilities, in order to infect \dev with malware.
  \item\label{sc:adv-model:la} \textit{Local Adversary} is sufficiently near \dev to intercept communication and generally interfere with network traffic between \sd and \dc or device-to-device communication between \dc and \dev.
  
\end{enumerate}
As in prior related literature, we consider attacks on \dev-s by so-called {\em physical adversaries} 
to be out-of-scope. However, we note that physical attacks can be mitigated via 
tamper-resistant techniques, or using communication-intensive (though unscalable) absence detection \cite{darpa}.

\subsection{Objectives}
\label{sec:our-goals}
We identify several objectives for a secure and reliable update framework applicable to realistic IoT devices.
These objectives also have some overlaps with the existing firmware update requirements discussed in~\cite{Moran-FUD1}.
\begin{itemize}	\item[\fbox{O1}] \textit{End-to-End Security:} \dev must verify that a firmware update it 
	receives is originated by \oem, and \oem must specify device-specific constraints on the update. However, 
	due to large numbers of devices, \oem may not be able to directly interact with all.
	\item[\fbox{O2}] \textit{Update Authorization from \dc:} \dc must control 
	which firmware updates must be installed on \dev. This implies that \dev must verify whether 
	firmware updates are approved by \dc for installation.
	\item[\fbox{O3}] \textit{Attestation of Update Installation:} \dc must obtain a 
	verifiable proof of successful update installation on \dev.
	\item[\fbox{O4}] \textit{Protection of Code \& Secret Keys on Device:} ensure confidentiality 
	and integrity of code and secret keys used in update and attestation processes.
		\item[\fbox{O5}] \textit{Minimal Burden for Device:} impose minimal computational and 
	storage burden on \dev.
		\end{itemize}
TUF and Uptane do not satisfy all of these requirements in realistic IoT scenarios.
In particular, 
TUF also requires \dev itself to make policy decisions about which updates to fetch and 
install, violating \fbox{O5}. In addition, several security features of TUF require multiple signature 
verifications using different keys, which makes TUF computationally expensive, further violating 
\fbox{O5}. 

Uptane overcomes these issues by introducing the director repository to provide update
decisions for each device and limits verification requirements for resource-constrained devices to only 
verifying director signatures. However, since the director repository is held by \oem, Uptane implies 
direct interaction between \oem and \dev, which violates \fbox{O1}. 
TUF and Uptane do not consider the client device as part of their threat models and 
simply assumes overall security of the device, not satisfying \fbox{O4}.
Lastly, neither of them specifies the need 
for an external entity to validate correct update installation, which violates \fbox{O3}.

\subsection{Device Prerequisites}
\label{sec:dev-req}
To meet aforementioned objectives, \dev's security architecture must include at least the following:
\begin{itemize}[$\bullet$]
  	\item \textbf{Secure or Authenticated Boot:} to guarantee authenticity and integrity of trusted software at boot time. 
	This generally requires a minimal hardware root-of-trust, e.g., as in~\cite{eldefrawy2012smart, 
	koeberl2014trustlite}. 		  	\item \textbf{Isolated Execution:} to protect trusted security-critical operations on \dev from being influenced 
	by untrusted (potentially vulnerable or malicious) code.
			  	\item \textbf{Secure Storage:} to ensure that trust anchors used for firmware update validation and 
	attestation are integrity-protected and only accessible by authorized trusted software at run-time.
		\end{itemize}
These requirements can be satisfied by modern embedded device platforms that support either 
(1) TrustZone Security Extensions~\cite{ARM-TrustZone}
or (2) a secure microkernel, e.g., seL4~\cite{klein2009sel4}. 
Section~\ref{sec:impl} discusses instantiations of our secure update framework on these two architectures.

\section{Design}
\label{sec:design}
\begin{figure*}[t]
  \centering
    \ifabridged
    \includegraphics[height=3.5in,width=\hsize]{figures/update}
    \else
    \includegraphics[width=\hsize]{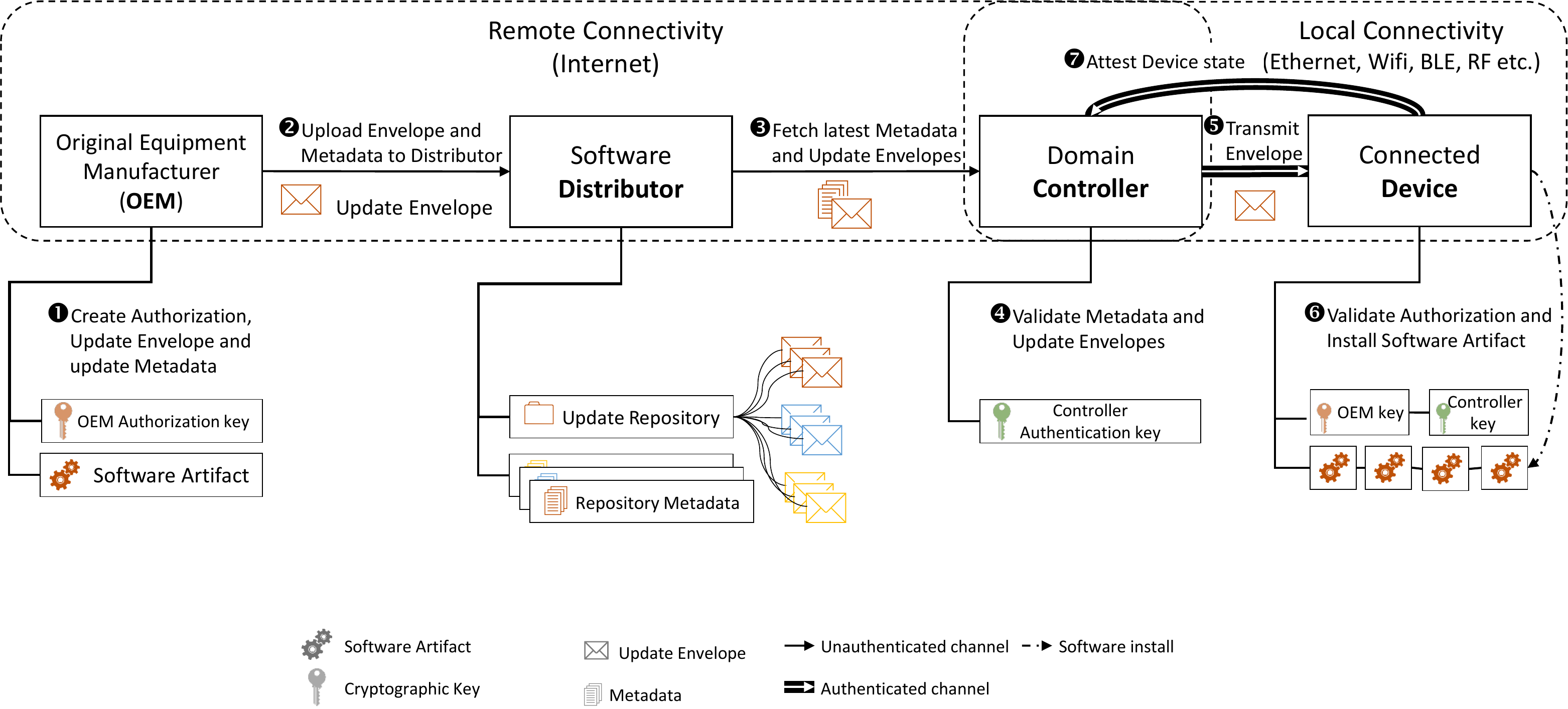}
    \fi
    \caption{Sequence of events during update distribution and delivery.} 
  \label{fig:update}
\end{figure*}

Our goal is to extend any update distribution scheme to allow \oem and \dc to specify constraints 
on the update process. As an example, we extend TUF with \acro and show how \dev can use \acro 
constraints to decide whether to install updates it receives. \acro can be combined seamlessly with 
TUF on \dc to benefit from TUF's security guarantees. As a result, besides security of TUF 
and Uptane, \acro satisfies additional \oem requirements on update distribution. 
We discuss how \acro satisfies the objectives in Section~\ref{sec:security}.

\acro expects \dev to implement the necessary mechanisms to meet \fbox{O1}, \fbox{O2}, and \fbox{O3}. 
However, \dev is not expected to perform full verification of TUF metadata. In Section~\ref{sec:impl}, we show that, 
as a result, \acro  compares favorably to TUF in terms of computational and storage burdens  on \dev. It is
thus very suitable for IoT deployments involving resource-constrained devices.

\subsection{Sequence of Events} 
\label{sec:soe}
\oem prepares \emph{Software Artifacts} for distribution by emitting cryptographic authorizations that can be 
verified by \dev to determine if software artifacts are sanctioned by \oem. An \emph{authorization token} 
encodes constraints in the form of metadata that is recognized by \dev, e.g., device model or a unique device 
identifier. This metadata must be validated by \dev when deciding if the software artifact should be installed. 
An authorization token must always include a signature computed with the \oem's authorization key on the 
hash of the constraints and the software artifact itself. 

Figure~\ref{fig:update} shows the sequence of events during update distribution and delivery. 
\oem emits an authorization token and encapsulates authorization information together with the 
corresponding software artifact in an \emph{Update Envelope} (\dOne). \oem uploads the resulting envelope
and its metadata to the \emph{TUF Repository} (\dTwo) where the envelope is recorded into the 
repository's TUF metadata. The TUF Repository is then mirrored by an untrusted \sd.

\dc, acting as a TUF client on behalf of \dev, periodically polls the repository for updates. 
When new update envelopes appear, \dc fetches the snapshot and targets metadata, 
validates them and fetches any new envelopes intended for \dev (\dThree). 
At this point, each envelope is validated against the corresponding record in the targets metadata (\dFour). 
\dc can now arbitrate on local update policies that may apply to the fetched software artifact. It then transmits the
update envelope (that it decides should be installed) to \dev over an authenticated channel (\dFive). 
This channel serves as an implicit authorization from \dc that it has approved the software artifact in the 
transmitted update envelope. \dev uses its underlying security architecture to securely validate authenticity 
and integrity of the OEM's authorization token and software artifact in the update envelope. If the signature and 
constraints are valid, it installs the software artifact (\dSix).
We note that the security architecture of \dev guarantees the protection of code and secret keys on \dev. 
Thus, \fbox{O4} is achieved in this step.
Finally, \dc attests the state of \dev to ensure that the software artifact is successfully installed (\dSeven).
The last step allows \dc to obtain a verifiable proof when the update process is complete, which satisfies \fbox{O3}.
Meanwhile, if the update process fails (e.g., by the adversarial preventing an update from reaching \dev),
\dc will be able to detect it due to the incorrect or missing response.

\subsection{Authorization Mechanism} 
\label{sec:auth-mechanism}
The mechanism for authorizing software updates must satisfy \fbox{O1} and \fbox{O2}. 
Namely, it must allow \dev to authenticate the source of software updates as well as let \oem and 
\dc specify applicable constraints. In \acro, we identify two concrete approaches for realizing 
authorization tokens that meet both needs:

\noindent\textbf{Extension of TUF Targets Metadata.} 
\oem can create an authorization token for each software artifact and embed it into the 
TUF targets metadata. This allows \oem to define update constraints for specific software artifact, 
as well as allows \dc to validate update metadata of different devices separately.
As a result, metadata associated with software artifact $SA$ in the targets metadata can now be encoded as:

\begin{center}
$\scriptstyle Auth_{SA}:=\left[hash\left(SA\right), size\left(SA\right), C, Sig\left(K_{OEM}, hash\left(SA\right) || size\left(SA\right) || C \right)\right] $
\end{center}
\noindent 
where $C$ denotes constraints, e.g., device model and/or unique device identifier. 

\noindent\textbf{Adoption of GP TMF.}
Alternatively, an authorization token can be delivered to \dev encapsulated in the Update Envelope, 
using GlobalPlatform TEE Management Framework (TMF)~\cite{GPD-TMF}. TMF is a security model for 
administration of TEEs. GlobalPlatform-compliant 
TEEs based on ARM TrustZone~\cite{ARM-TrustZone} are widely deployed, especially on 
Android devices. TMF defines the set of administration operations available to various parties 
in the administration of a TEE and its Trusted Applications (TAs). TMF also defines a security 
model that allows business relationships and responsibilities to be mapped to a set of 
\emph{Security Domains}, and a security layer for the authentication and establishment of secure 
communication channels between such parties and corresponding security domains.

The subset of TMF needed to support \acro authorization is only 
the \emph{Update TA} command \cite[Section 8.4.3]{GPD-TMF} and the use of TMF's explicit 
authorization~\cite[Section 5.2.1]{GPD-TMF} primitives. Explicit authorization allows 
TMF commands to be authenticated when there is no means of establishing a direct 
communication channel between the party that signs the authorization, and the on-device 
security domain acting on behalf of that party, such as in the case of broadcast channels 
and  update repositories. In the context of our framework, \oem signs a TMF Authorization Token
with associated constraints (such as the applicable device model) and emits a TMF Envelope that 
encapsulates the Software Artifact, Update TA command, and TMF Authorization Token. 

Both approaches could be realized in either TrustZone-M and HYDRA architectures.
However, since TrustZone-M is likely to be found on low-end MCUs, 
implementations of \acro based on concise binary encoding of update metadata, 
such as \emph{Concise Binary Object Representation}~\cite{RFC7049} or \emph{Abstract Syntax Notation One} (ASN.1)
are more suitable for TrustZone-M devices compared to TUF JSON objects. TMF is based on a 
subset of the ASN.1 Distinguished Encoding Rules, and in addition provides an existing set of 
constraints that can be easily extended~\cite[Section 5.3.2]{GPD-TMF}. Alternatively, \oem or \dc 
can encode $Auth_{SA}$ in fixed-size formats suitable for parsing on severely restricted devices.

\section{Implementation}
\label{sec:impl}
In this section we describe two proof-of-concept implementations of \acro.

\begin{figure}[!t]
  \centering
	 \includegraphics[width=1.1\linewidth]{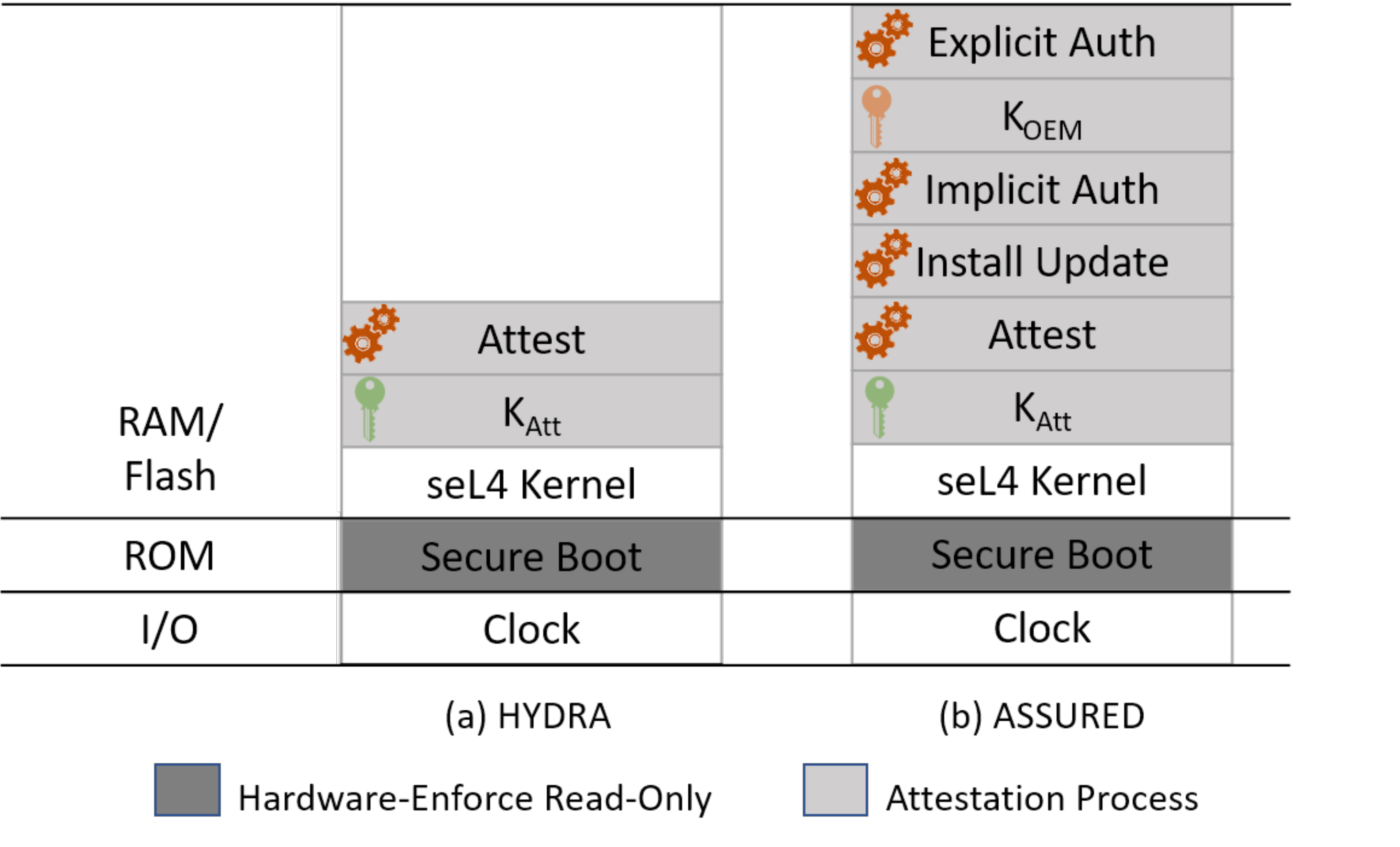}
	\caption{Memory organization of \hydra-based firmware update.}
	\label{fig:hydra}
\end{figure}

\subsection{\acrotitle on HYDRA}
\label{sec:assured-hydra}
We now overview \hydra, discuss implementation details and report on experimental evaluation.

\subsubsection{HYDRA Overview}
\hydra implements a hybrid (HW/SW) remote attestation design by building upon the 
formally verified seL4 \cite{klein2009sel4} microkernel, which provably guarantees process memory 
isolation and enforces access control to memory regions. 
Using the formally proven isolation features of seL4, access control rules can be 
implemented in software and enforced by the microkernel. 
Figure~\ref{fig:hydra}a summarizes memory organization of \hydra.
\hydra implements secure storage for the attestation key (\attkey) by storing it in a writable memory region and configuring the system, such that no other 
process, besides the attestation process (\patt), can access this memory region. 
Access control configuration in \hydra guarantees strong isolated execution of \patt by enforcing 
exclusive access to its thread control block as well as to its memory regions. 
To ensure uninterruptibility, \hydra runs \patt as the so-called {\em initial user-space process} 
with the highest scheduling priority.  As the initial user-space process in seL4, \patt is initialized with 
capabilities that allow access to all available resources.

Meanwhile, the rest of user-space processes are assigned lower priorities and their resource access 
is limited by \patt. \hydra also requires a {\em reliable read-only clock} to defend against denial-of-service 
attacks via replayed, delayed or reordered attestation requests \cite{brasser2016remote}. 
Finally, hardware-enforced secure boot feature is used to ensure integrity of seL4 itself and of 
the initial process when the system is initialized.

\subsubsection{Implementation Details}
Figure~\ref{fig:hydra}b shows the implementation of \acro as part of \patt in \hydra.
Specifically, we modify \patt to support the TMF-style authentication mechanism 
via implicit and explicit authorization operations.

Following TMF specifications~\cite{GPD-TMFSCSL}, we use AES~\cite{aes} and 
HMAC-SHA256~\cite{hmac-sha256} as the underlying cryptographic 
primitives to ensure implicit authorization from \dc via a secure channel. In particular, \patt derives 
encryption and MAC keys (used during the setup of the secure channel) from a pre-shared master \attkey.
Once established, the secure channel between \patt and \dc yields new session keys 
used to protect the transmitted update envelope.
 
For explicit authorization, we assume \oem's authorization key (\oemkey) is distributed and pre-installed on 
devices out-of-band, e.g., during manufacturing. To ensure confidentiality and integrity, \patt protects \oemkey 
the same way as \attkey. 
Elliptic Curve-based signature scheme, ED25519~\cite{bernstein2012high}\footnote{ED25519 is chosen because
it is shown to run faster than other existing signature schemes while still providing the same security guarantees.}, is ported 
to seL4 and serves as the underlying signature scheme to provide explicit authorization operation in \patt.

\begin{table*}[htb]
\centering
\resizebox{\columnwidth}{!}{\begin{tabular}{lcccc}
\dtoprule 
	& \multirow{2}[3]{*}{TUF} & \multicolumn{3}{c}{\sf ASSURED}\\
	\cmidrule(lr){3-5}
    	& & Expl. Auth. & Impl. Auth. & Total \\
    	\midrule
    	Verification Time (ms)   & 14.57 & 2.46 & 0.1  & 2.56 \\
    	Metadata Size (bytes)  & 940  & 136  & 52 & 188 \\
    	\bottomrule
\end{tabular}
}
\caption{Performance comparison between \acro and TUF on I.MX-SabreLite @ 800MHz. Verification of 
TUF metadata is performed using TUF-recommended threshold values: 2 for root and targets roles, 
and 1 for others. TUF metadata size is estimated assuming only one target file in the targets role.}
\label{tab:verf-time}
\end{table*}

\begin{figure}[!t]
  \centering
	 \includegraphics[width=\columnwidth]{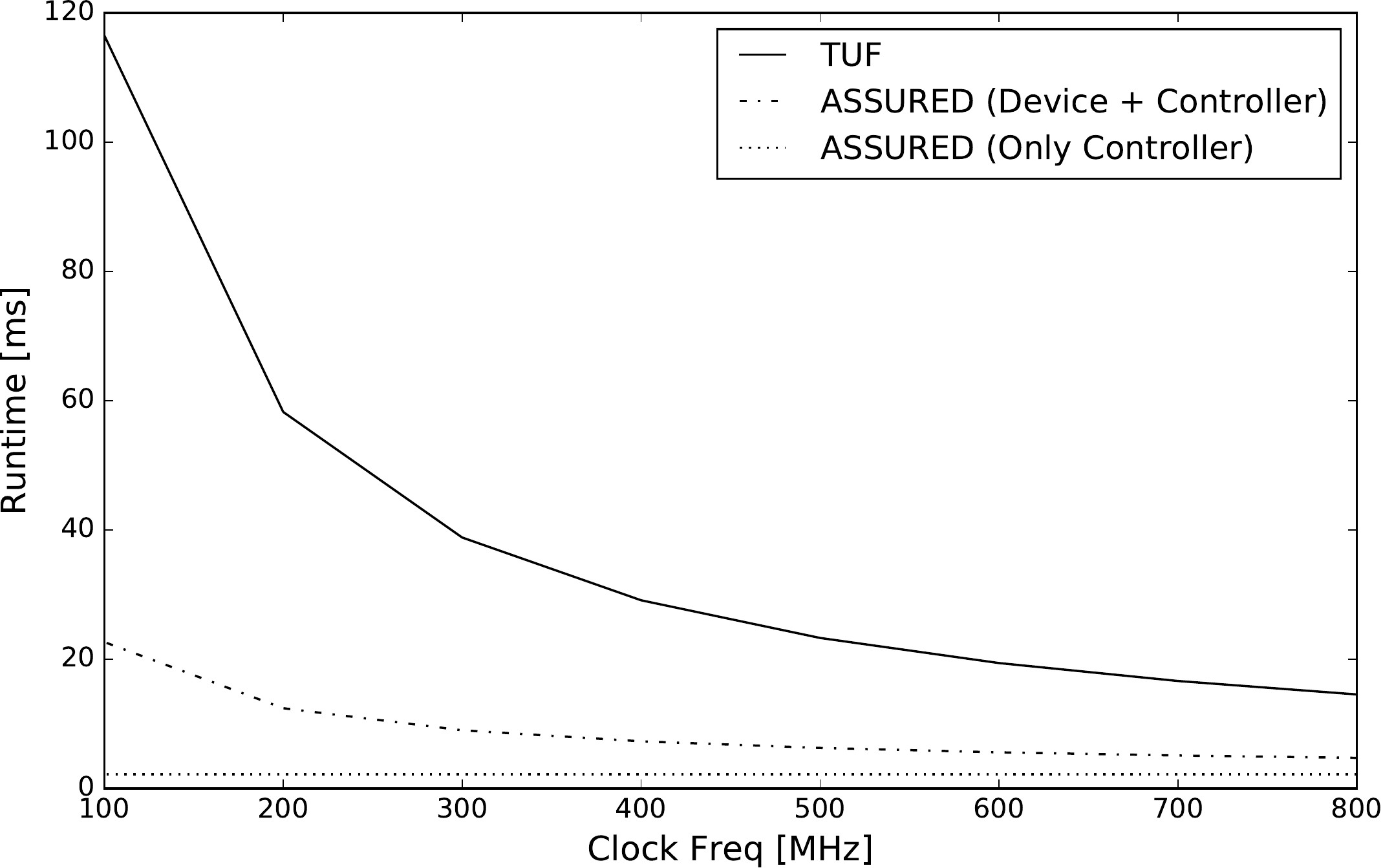}
	\caption{Total runtime of TUF and \acro for variable \dev clock frequencies (I.MX6-SabreLite). 	
	\label{fig:runtime-lowfreq}	
	\dc clock frequency is fixed at 3.4GHz.}
\end{figure}

\begin{table*}[htb]
\centering
\begin{tabular}{lcccc}
	\dtoprule
	&  \multicolumn{2}{c}{\sf HYDRA} & \multirow{2}[3]{*}{\sf ASSURED} & \multirow{2}[3]{*}{ED25519 Impl.}  \\
	\cmidrule(lr){2-3}
	& Attestation & Net. \& Libs. & \\
    	\midrule
    	Code Size (KLOC)   & 12 & 94 & 2.2 & 4.5  \\
    	Executable Size (KB)   & \multicolumn{2}{c}{---------  250.6  ---------} & \multicolumn{2}{c}{----------- 9.4  -----------}  \\
	\bottomrule
\end{tabular}
\caption{Code and executable sizes of \patt on I.MX6-SabreLite.}
\label{tab:code-size}
\end{table*}

\subsubsection{Evaluation}
\label{sec:eval}
We compare performance of \acro and TUF in terms of code size and runtime on a popular 
commercially available platform: I.MX6-SabreLite~\cite{sabre}. We chose TUF as a point of 
comparison since it (and its variants) is currently the only established secure update 
standard relevant to IoT~\cite{uptane-news}.

\noindent\textbf{Code \& Metadata Size.}
As shown in Table~\ref{tab:code-size}, \acro adds around $6.7K$ lines of C 
code to \hydra's code-base, while overall size of \patt executable increases by $9$KB.
About 67\% of code overhead is due to ED25519 code. In order to minimize runtime from 
parsing metadata, we encode \acro's metadata in a fixed-size format instead of JSON.
An \acro update envelope carries $188$ bytes of metadata,
while the size of TUF metadata is estimated to be around $940$ bytes. 
(See Table~\ref{tab:verf-time} for more details.)

\noindent\textbf{Runtime Overhead.}
Table~\ref{tab:verf-time} shows the runtime comparison between \acro and TUF implemented on top of \hydra.
Recall that we use ED25519~\cite{bernstein2012high} as the signature scheme for both methods.
In a typical scenario, full verification of all TUF metadata takes much longer ($\sim~5.7$ times) on \dev, 
for two reasons. First, since TUF metadata is encoded in JSON format,
parsing it on \dev as part of the update process consumes a non-negligible amount of time.
In our experiments, this takes around $1ms$ or $\sim~6\%$ of total runtime. 
However, the major reason for this significant increase is because TUF full verification requires 
at least $6$ public key operations. In contrast, \acro offloads these operations to \dc, and \dev only performs
lighter-weight computation, i.e., validating an \oem authorization token received from \dc via an authenticated channel.
Figure~\ref{fig:runtime-lowfreq} shows that \dc performs TUF full verification in $2.2ms$.

Next, we assess runtime performance of the entire \acro process, i.e., combined runtime of \acro in both \dev and \dc,
and compare it to runtime for \dev to perform full TUF verification. Results in Figure~\ref{fig:runtime-lowfreq} show 
that \acro is still considerably faster than TUF and the difference becomes more significant as \dev's clock frequency drops.
This clearly serves as a motivation to offload this computationally expensive task to \dc.

\subsection{\acrotitle on ARM Cortex-M23}
\label{sec:assured-tzm}
We now describe a proof-of-concept implementation of \acro on a 
Cortex-M23 MCU and report on its experimental evaluation.

\subsubsection{ARM Cortex-M23 Overview}
As described in Section~\ref{sec:trustzone} ARM Cortex-M23 MCU is equipped with TrustZone Security Extensions 
that allow partitioning the system into \emph{secure} (trusted) and \emph{non-secure} (untrusted) execution environments.
(Sometimes these are referred to as separate secure and non-secure \emph{"worlds"})  separated from each other by hardware.) 
A context switch between them is performed by the hardware processor logic when specific conditions are met. The processor logic 
ensures that code in the non-trusted execution environments can enter trusted code only at specific entry points, and that non-trusted 
code remains strongly isolated from resources (e.g., memory and interrupt lines) belonging to the trusted execution environment. 

At system boot, the MCU starts execution in the trusted execution environment. Although the boot flow might vary between 
specific Systems-on-Chip (SoCs), it typically begins from bootstrap code stored in secure ROM that validates and starts a 
trusted bootloader, e.g., based on a trust root for verification, often reflected in the hash of a code signature verification key stored in 
one-time-programmable fuses. 

The trusted bootloader configures access control rules for memory partitioning to separate trusted code and data from their non-trusted 
counterparts. Secure storage can be realized by simply storing sensitive keys (or other data) in memory allocated to the trusted execution 
environment. The trusted bootloader can also adjust interrupt priorities and interrupt line assignments to ensure that trusted code 
receives priority when deciding which interrupt handler routines are invoked to service processor events.

To ensure that non-trusted code can not change device's software configuration, persistent storage used as code memory 
(e.g., internal flash) can be configured to be only writable by trusted code. Code re-programming support may be exposed 
to non-trusted code via APIs provided by trusted software. These APIs can implement authentications to decide if re-programming 
of code is allowed.
\begin{table*}[hbt]
\centering
\resizebox{\columnwidth}{!}{\begin{tabular}{lcc}
\dtoprule
	& TUF & {\sf ASSURED}\\\hline
    	Verification Time (ms)         & 10723 & 1816 \\
	Attestation Response Time (ms) &   N/A & 517 \\
    	Metadata Size (bytes)   &   940 &  136 \\
    	\bottomrule
\end{tabular}
}
\caption{Performance comparison between \acro and TUF on ARM V2M-MPS2+ configured as a Cortex-M23 @ 25MHz. 
Verification of TUF metadata is performed using the same parameters as in Table~\ref{tab:verf-time}.}
\label{tab:verf-time-tzm}
\end{table*}

\begin{table*}[hbt]
\centering
\resizebox{\columnwidth}{!}{\begin{tabular}{lccc}
	\dtoprule
	& Bootloader & {\sf ASSURED} & ED25519 Impl. \\
    	\midrule
    	Code Size (LoC)  & 959 & 5448 & 4322  \\
	Executable Size (KB)  & \multicolumn{2}{c}{------ 48 ------} & ~17 \\
	\bottomrule
\end{tabular}
}
\caption{Code and executable sizes of \patt on ARM V2M-MPS2+.}
\label{tab:code-size-tzm}
\end{table*}

\subsubsection{Implementation Details}
We implemented \acro as part of the trusted bootloader on \patt. In this variant, we only support explicit authorization 
via a pre-configured trust root for verification in the form of a public authorization key (\oemkey), which is embedded 
into the trusted bootloader software image placed on \dev during manufacturing.  $Auth_{SA}$ is encoded in a 
fixed-size format. As in \acro on \hydra, we use ED25519 as the signature scheme.

\attkey is established with \dc during enrollment and stored in secure memory.
$Auth_{SA}$ is stored in persistent storage on \patt, and used during \dev boot to validate non-trusted software artifacts ($SA$s). 
If \dev has sufficient memory to store both the current software artifact $SA_{n}$ and the next update $SA_{n+1}$, 
$Auth_{SA_{n+1}}$ is validated before $SA_{n}$ is reprogrammed with $SA_{n+1}$. However, if \dev can not store 
both $SA_{n}$ and $SA_{n+1}$ simultaneously, $SA_{n}$ is overwritten by $SA_{n+1}$ and only then validated. 
In the latter case, if $SA_{n+1}$ validation fails, a replacement $SA$ must be obtained from \dc. We recommend that a 
back-up copy of the trusted bootloader is kept when updating the trusted bootloader itself to ensure that the 
update process remains robust. Hence, it is important to minimize the impact of \acro on the trusted bootloader code size.

\subsubsection{Evaluation}
We assess performance -- in terms of code and size and runtime -- of \acro for resource-constrained 
MCUs on ARM Versatile Express Cortex-M Prototyping System MPS2+ FPGA 
(ARM V2M-MPS2+)\footnote{\url{https://www.keil.com/boards2/arm/v2m\_mps2/}} configured as a 
Cortex-M23 MCU running at 25~MHz. 

\noindent\textbf{Code \& Metadata Sizes.}
Table~\ref{tab:code-size-tzm} shows the impact of \acro on trusted bootloader code size. Most of the increase in code size is 
attributed to the ED25519 implementation -- $\approx~80\%$. The size of $Auth_{SA}$ is a mere $136$ bytes.

\noindent\textbf{Runtime Overhead.} 
To compare \acro with TUF, we adapted the TUF implementation from Section~\ref{sec:assured-hydra} to run on 
ARM Cortex-M23 MCU. As before, we used ED25519 as the signature scheme for both \acro and TUF. 
TUF uses a fixed-size encoding for its metadata. Table~\ref{tab:verf-time-tzm} shows the runtime comparison 
between \acro and TUF on Cortex-M23. Our assessment of runtime performance of \acro includes 
validation $Auth_{SA}$ and $SA$ on \dev, compared with full TUF verification. We also measured the 
time for \dev\ to generate its attestation response. 

The evaluation shows that \acro outperforms TUF (when using full metadata 
verification) by a factor of ~$4.5$ in terms of total time spent for metadata 
verification and attestation response generation.

\section{Meeting Stated Objectives}
\label{sec:security}
We use descriptions of \acro design and realization (in Sections~\ref{sec:design} 
and~\ref{sec:impl}, respectively) to informally argue that \acro satisfies all objectives stated 
in Section~\ref{sec:our-goals}.
\begin{itemize}
	\item[\fbox{O1}] \textit{End-to-End Security:} 
	\acro requires \oem to include an authorization token in each update envelope. 
	End-to-end security with constraints between \oem and \dev is thus guaranteed, since the token represents 
	explicit authorization from \oem, which can be validated by \dev without the need to establish 
	a direct communication channel with \oem.
	\item[\fbox{O2}] \textit{Update Authorization from Controller:} 
	\acro requires \dc to transmit an update envelope to \dev through an authenticated channel. 
	This serves as implicit authorization by \dc that it has approved the software artifact contained in the envelope.
	\item[\fbox{O3}] \textit{Attestation of Update Installation:} 
	At the end of \acro's sequence of events, \dev must reply to \dc with an attestation result that reflects  
	its current software state. This allows \dc to determine whether the update has been correctly installed on \dev.
	\item[\fbox{O4}] \textit{Protection of Code \& Secret Keys on \dev:} 
		The underlying \hydra architecture provides secure storage for 
	secret keys using capability-based access control configuration, and isolated execution of 
	critical code guaranteed by seL4.  Also, our specific hardware platform (SabreLite) provides 
	hardware-enforced secure boot of seL4. In ARM Cortex-M23, this property is satisfied similarly by: 
	(1) TrustZone Security Extensions that allow partitioning for a secure environment
	and (2) a secure boot chain anchored in ROM-resident bootstrap code.
	Therefore, both implementation of \acro (on \hydra and ARM Cortex-M23) satisfy
	all \dev requirements and meets this objective.
	\item[\fbox{O5}] \textit{Minimal Burden for \dev:} 
	As experimental results show, \acro considerably lowers 
	computational burden on \dev, by off-loading heavy computational tasks to \dc.
	However, we do not claim that the incurred overhead is truly minimal.
\end{itemize}

\section{Conclusion}
This paper motivates the need for, and constructs \acro\ -- a secure firmware update framework 
for the large-scale IoT setting with resource-constrained devices. \acro extends TUF -- the popular 
state-of-the-art secure update mechanism. \acro takes into account realistic stakeholders in 
large-scale IoT deployments while providing end-to-end security with enforceable constraints 
between device manufacturers and IoT devices.
\acro offloads heavy computational operations to more powerful entities and places
minimal burden on IoT devices. Practicality of \acro is demonstrated via two prototype 
implementations on 
\hydra and ARM TrustZone-M architectures. Experimental evaluations show that \acro
incurs very low overhead, particularly for end-devices.

\section*{Acknowledgments}

Research by UCI co-authors was supported in part by: (1) DHS, under subcontract from HRL Laboratories, (2) ARO under contract W911NF-16-1-0536, and (3) NSF WiFiUS Program Award 1702911. Research by Aalto University co-authors was supported in part by (1) Academy of Finland under grant nr. 309994 (SELIoT) under the auspices of the WiFiUS program, (2) Business Finland under grant nr. 3881/31/2016 (CloSer), and (3) the Intel Collaborative Research Institute for Collaborative Autonomous \& Resilient Systems (CARS). Research by TU Darmstadt co-authors was supported in part by German Science Foundation CRC 1119 CROSSING project S2.

\ifabridged
\bibliographystyle{IEEEtranS}
\else
\bibliographystyle{ACM-Reference-Format}
\fi
\bibliography{sw-update} 

\end{document}